\documentclass[twocolumn,preprintnumbers,nofootinbib]{revtex4}
\usepackage{graphicx}
\usepackage{amssymb}
\usepackage{color}
\usepackage{enumerate}
\def\beq{\begin{equation}}
\def\eeq{\end{equation}}
\def\beqn{\begin{eqnarray}}
\def\eeqn{\end{eqnarray}}

\renewcommand{\texttt}{{}}
\newcommand{\be}{\begin{eqnarray}}
\newcommand{\ee}{\end{eqnarray}}

\begin{document}

%
%
\title{The Hidden Quantum Groups Symmetry of Super-renormalizable Gravity}
%
%
%

\author{Stephon Alexander}
\email{salexand@haverford.edu}
\affiliation{Department of Physics, The Koshland Integrated Natural Science Center, Haverford College, Haverford, PA 19041 USA}
\affiliation{Department of Physics, Princeton University, New Jersey 08544, USA}
\affiliation{Institute for Gravitation and the Cosmos, Penn State University, University Park, PA, 16802}

\author{Antonino Marcian\`o}
\email{amarcian@haverford.edu}
\affiliation{Department of Physics, The Koshland Integrated Natural Science Center, Haverford College, Haverford, PA 19041 USA} 
\affiliation{Department of Physics, Princeton University, New Jersey 08544, USA}

\author{Leonardo Modesto}
\email{lmodesto@perimeterinstitute.ca}
\affiliation{Perimeter Institute for Theoretical Physics, 31 Caroline St., Waterloo, ON N2L 2Y5, Canada}

\date{\small\today}

\begin{abstract} \noindent
In this paper we consider the relation between the super-renormalizable theories of quantum gravity (SRQG) studied in \cite{BM, Modesto} and an underlying non-commutativity of spacetime. For one particular super-renormalizable theory we show that at linear level (quadratic in the Lagrangian) 
the propagator of the theory is the same we obtain starting from a theory of gravity endowed with $\theta$-Poincar\'e quantum groups of symmetry. Such a theory is over the so called $\theta$-Minkowski non-commuative spacetime. We shed new light on this link and show that among the theories considered in \cite{BM, Modesto}, there exist only one non-local and Lorentz invariant super-renormalizable theory of quantum gravity that can be described in terms of a quantum group symmetry structure. 
We also emphasize contact with pre-existent works in the literature and discuss preservation of the equivalence principle in our framework.

\end{abstract}
\pacs{05.45.Df, 04.60.Pp}
\keywords{perturbative quantum gravity, nonlocal field theory}

\maketitle

\section{Introduction} 


\noindent In the recent papers \cite{BM, Modesto} it has been introduced a modified theory of gravity assuming a synthesis of minimal requirements: (i) regularity of classical solutions;
(ii) Einstein-Hilbert action should be the correct low energy limit;
(iii) the spacetime dimension has to decrease with the energy; 
(iv) the theory has to be perturbatively renormalizable at quantum level;  
(v) the theory has to be unitary, with no other 
pole beyond the graviton in the propagator.

The theory 
we are going to summarize in the next section is power counting super-renormalizable 
at the quantum level at least perturbativelly and at classical level the gravitational potential \cite{BM}, black hole solutions 
\cite{Modesto, NS1, NS2, NS3, NS4, NS5, NS6, NS7, NS8, NS9, NS10, NS11, NS12, NS13, NS14, NS15, NS16, NS17, NS18, NS19, NS20, NS21, ModestoMoffatNico, Unpart, NL1} and the cosmological solutions are singularity free \cite{BM, NL2, NL3}. The Lagrangian is a ``nonlocal" extension of the renormalizable quadratic Stelle theory \cite{Stelle} but the non locality only involve positive powers of the D'Alembertian covariant operator. 
In other words there are not operators like $1/\Box^p$ ($p>0$). 
The theory is not unique (we thus refer to super-renormalizable ``theories''), but all the freedom present in the action can be read in an ``entire function" of the D' Alembertian operator, $H( - \Box/\Lambda^2)$ \cite{Tombo} ($\Lambda$ is a physical mass-invariant scale introduced in the classical action). 

 The reason of this paper is not only to find an elegant reason for the nonlocal nature of the action, but it is to find a way to fix uniquely the entire function which is mentioned above. In this paper we show that the propagator of the theory, for a particular choice of the entire function $H( - \Box/\Lambda^2)$, has exactly the same form of  the propagator we obtain starting from a theory of gravity endowed with $\theta$-Poincar\'e quantum groups of symmetry. The right choice is much easier we could think, i.e. $H( - \Box/\Lambda^2) =  - \Box/\Lambda^2$. Any other entire function gives of course a well defined super-renormalizable theory of gravity (consistently with some particular properties \cite{BM, Modesto}) but is not compatible with the requirement of having a non-trivial Hopf-algebra-like symmetry regulating the super-renormalizability of the theory. In particular, the Hopf-algebra underlying the super-renormalizable model we discuss below is a quantum-group associated to an associative non-commutative space-time. In particular, this is the only quantum group of (space-time) symmetry that can be accounted within the model presented in \cite{Modesto}, if we do not relax the associativity of the space-time points' coordinates. What emerges is therefore a new symmetric structure underlying the theory.

\section{The theory} \label{due}

\noindent A simplified version of the theory is a nonlocal generalization of the Stelle quadratic action for gravity \cite{Stelle} and can be written 
in the following compact form, 
\be 
&& \hspace{-0.3cm}
\mathcal{L}_g =   - \sqrt{ - g} \, \left[ \frac{\beta}{\kappa^2} R
+ R_{\mu \nu}  \, F(-\Box_{\Lambda})^{\mu\nu \rho \sigma} R_{\rho \sigma} \right], \label{theory} 
\ee
where the tensor $F( - \Box_{\Lambda})$ is a function of the covariant D'Alembertian operator $- \Box_{\Lambda} := - \Box/\Lambda^2$, $\Lambda$ is a physical mass scale and $\kappa^2=32 \pi G$. To fix the notation we can write more explicitly the tensor $F( - \Box_{\Lambda})$
in terms of two entire functions $h_2$ and $h_0$ that we are going to define in this same section,
\be
&& F(-\Box_{\Lambda})^{\mu\nu \rho \sigma} := 
- \Big( \beta_2 - h_2( - \Box_{\Lambda} ) \Big) g^{\mu \rho} g^{\nu \sigma} \nonumber \\
&& + \Bigg(  \frac{\beta_2}{3} + \beta_0   -   h_0 ( - \Box_{\Lambda}) 
- \frac{h_2( - \Box_{\Lambda}  ) }{3}  \Bigg)  g^{\mu \nu} g^{\rho \sigma}.
\ee
The complete Lagrangian including also the gauge fixing and ghost terms is 
\be
\mathcal{L} = \mathcal{L}_g +  \mathcal{L}_{\rm GF, GH}, 
\ee
where the gauge fixing and ghost Lagrangian terms are
\be
 \mathcal{L}_{\rm GF, GH} = -\frac{1}{2 \xi} F^{\mu}  \omega( - \Box_{\Lambda}^{\eta})  F_{\mu} + 
\bar{C}^{\mu} M_{\mu \nu} C^{\nu} \label{gaugec}.
\ee
The operator $\Box^{\eta}_{\Lambda}$ encapsulates the D'Alembertian of the flat fixed background, whereas $F_{\mu}$ is the gauge fixing function with the weight functional $\omega$. The two functions $h_2$ and $h_0$ have not to be polynomial but ``entire functions without poles or essential singularities" to avoid ghosts (states with negative norm) in the spectrum $\omega$. $\bar{C}^{\mu}, C^{\mu}$ are the ghosts fields and $M^{\tau}_{\,\,\alpha} = F^{\tau}_{\,\,\,\mu \nu} D^{\,\,\,\mu \nu}_{\alpha}$. The gauge fixing function $F^{\tau}_{\,\,\,\mu \nu}$ and the operator $D^{\,\,\,\mu \nu}_{\alpha}$ will be defined shortly, in (\ref{daddi}).

We calculate now the graviton propagator. For this purpose we start by considering the quadratic expansion of the Lagrangian (\ref{theory}) in the graviton field fluctuation without specifying the explicit form of the functionals $h_2$ and $h_0$ (if not necessary). Following the Stelle paper \cite{Stelle}, we expand around the Minkowski background $\eta^{\mu \nu}$ in power of the graviton field $h^{\mu \nu}$  defined in the following way  
\begin{eqnarray} \label{peba}
\sqrt{ - g} g^{\mu \nu} = \eta^{\mu \nu} + \kappa h^{\mu \nu}. 
\label{graviton}
\end{eqnarray}
The form of the propagator depends not only on the gauge choice but also on the definition of the gravitational fluctuation \cite{Shapirobook}. The gauge choice is the familiar ``harmonic gauge" $\partial_{\nu} h^{\mu \nu} =0$ and in (\ref{gaugec}), $F^{\tau} = F^{\tau}_{\mu \nu} h^{\mu \nu}$ with $F^{\tau}_{\mu \nu} = \delta^{\tau}_{\mu}\partial_{\nu}$.  $D^{\,\,\,\mu \nu}_{\alpha}$ is the operator which generates the gauge transformations in the graviton fluctuation $h^{\mu \nu}$. Given the infinitesimal coordinates transformation $x^{\mu \prime} = x^{\mu} + \kappa \xi^{\mu},$ the graviton field transforms as follows 
\be \label{daddi}
&& \delta h^{\mu \nu} = D^{\,\,\,\mu \nu}_{\alpha} \xi^{\alpha} 
= \partial^{\mu} \xi^{\nu} + \partial^{\nu} \xi^{\mu} - \eta^{\mu \nu}  \partial_{\alpha} \xi^{\alpha} \nonumber \\
&& \hspace{-0.5cm}
+ \kappa ( \partial_{\alpha} \xi^{\mu} h^{\alpha \nu} + \partial_{\alpha} \xi^{\nu} h^{\alpha \mu}
- \xi^{\alpha} \partial_{\alpha} h^{\mu \nu} - \partial_{\alpha} \xi^{\alpha} h^{\mu \nu} ).
\ee

We Taylor-expand now the gravitational part of the action (\ref{theory}) to the second order in the gravitational perturbation $h^{\mu \nu}(x)$ to obtain the graviton propagator. In the momentum space, the action which is purely quadratic in the gravitational field, reads  
\begin{eqnarray}
\mathcal{L}^{(2)} = \frac{1}{4} 
h^{\mu \nu} (-k) 
 K^{\mu \nu \rho \sigma} h^{\rho \sigma}(k) + \mathcal{L}_{\rm GF}, 
\label{quadratic}
 \end{eqnarray}
where $\mathcal{L}_{\rm GF}$ is the gauge fixing Lagrangian at the second order in the graviton field 
\be
&&  \hspace{-1.1cm}
\mathcal{L}_{\rm GF} = \frac{1}{4 \xi}
 h^{\mu \nu} (-k) 
\Big( 
\,  \omega(k^2/\Lambda^2)  k^2 P^{(1)}_{\mu \nu \rho \sigma}(k) \nonumber \\
&& + 2  \omega(k^2/\Lambda^2) k^2 \} P^{(0 - \omega)}_{\mu \nu \rho \sigma}(k) \Big)
 h^{\rho \sigma}(k).
\label{GF}
\ee
The kinetic operator $K_{\mu \nu \rho \sigma}$ is defined by 
\begin{eqnarray}
&& \hspace{-0.4cm} 
K_{\mu \nu \rho \sigma} := 
  -  \bar{h}_{2}(z)    \, k^2 \, P^{(2)}_{\mu \nu \rho \sigma}(k)
+  \frac{3}{2} \, k^2 \,   \bar{h}_0(z)  \, 
P^{(0 - \omega)}_{\mu \nu \rho \sigma}(k) \nonumber \\
&& \hspace{-0.5cm}
+ \frac{k^2}{2} \,  \bar{h}_0(z) \, 
\{   P^{(0 - s)}_{\mu \nu \rho \sigma}(k) + \sqrt{3} [ P^{(0 - \omega s)}_{\mu \nu \rho \sigma}(k)
+ P^{(0 - s \omega)}_{\mu \nu \rho \sigma}(k) ] \}  , \nonumber 
\end{eqnarray}
and we have introduced the following notation 
\be
&& \bar{h}_2(z) := \beta - \beta_2 \kappa^2 \Lambda^2 z + \kappa^2 \Lambda^2 z \, h_2(z) , \nonumber \\
&& \bar{h}_0(z) := \beta - 6 \beta_0 \kappa^2 \Lambda^2 z + 6 \kappa^2 \Lambda^2 z \, h_0(z), 
\label{hbar}
\ee
where $z := - \Box_{\Lambda}$. Notice that in (\ref{quadratic}) $\Box_\Lambda$ has to be identified with the D'Alembertian operator in flat spacetime $ - \Box^{\eta}_{\Lambda}$. We have used the gauge $F^{\tau}= \partial_{\mu} h^{\mu \tau}$ and introduced the projectors $P^{(2)}, P^{(1)},  P^{(0-s)},  P^{(0-s\omega)}, P^{(0-\omega s)}$ \cite{VN} (see also appendix \ref{P}). Using the orthogonality properties of the projectors we can now invert the kinetic matrix in (\ref{quadratic}) and obtain the graviton propagator. In the following expression the graviton propagator is expressed in the momentum space according to the quadratic Lagrangian (\ref{quadratic}), 
\be
D_{\mu\nu\rho\sigma}(k) = D^{\xi=0}_{\mu\nu\rho\sigma}(k) + D^{\xi}_{\mu\nu\rho\sigma}(k),
\label{fullprop}
\ee
where the propagator in the gauge $\xi = 0$ is 
\be
\hspace{0.2cm} D^{\xi=0}_{\mu\nu\rho\sigma}(k) = \frac{- i}{(2 \pi)^4} \frac{2}{k^2 + i \epsilon} 
\Bigg( \frac{P^{(2)}_{\mu \nu \rho \sigma}(k)}{\bar{h}_2(k^2/\Lambda^2)}
- \frac{2 P^{(0 - s)}_{\mu \nu \rho \sigma}(k) }{ \bar{h}_0(k^2/\Lambda^2)} \Bigg) 
\nonumber 
\ee
and $D^{\xi}_{\mu\nu\rho\sigma}(k)$ is the gauge dependent part of the propagator. 

We are now in the position to find an upper bound to the divergences in quantum gravity. 
We consider a particular theory in which the two general entire functions $h_i(z)$ 
introduced in the action have the following asymptotic exponential behavior,
\be
&& h_2(z) = \frac{\alpha(e^z - 1) + \alpha_2 z}{\kappa^2 \Lambda^2 z}, \nonumber \\
&& h_0(z) = \frac{\alpha(e^z - 1) + \alpha_0 z}{6 \kappa^2 \Lambda^2 z},
\label{hz}
\ee
for three general parameters $\alpha$, $\alpha_2$ and $\alpha_0$.

Given the ultraviolet exponential behavior of the two functions ${h}_i(z)$, let us study the high energy behavior of the quantum theory. The ultraviolet behavior of the propagator in momentum space  (actually we will see that this is the correct scaling of the propagator at any energy scale), omitting the tensorial structure, reads 
\be
D(k) \sim \frac{e^{- k^2/\Lambda^2}}{k^{2}}\,.
\ee
But also the $n$-graviton interaction has the same scaling in the momentum space, since it can be written in the following schematic way
\be
&& \hspace{-0.25cm}
{\mathcal L}^{(n)} \sim  h^n \, \Box_{\eta} h \,\,  h_i( - \Box_{\Lambda}) \,\, \Box_{\eta} h \nonumber \\
&& \hspace{0.5cm}
\rightarrow h^n \, \Box_{\eta} h 
\,  \frac{e^{- \Box_{\eta} }}{\Box_{\eta}} \, 
\Box_{\eta} h + \dots \, , 
\label{intera}
\ee
in which ``$\dots$" indicates other interaction terms 
coming from the
covariant D'Alembertian and $\Box_{\eta} = \eta^{\mu \nu} \partial_{\mu} \partial_{\nu}$.

Placing an upper bound to the amplitude with $L$-loops, we find 
\begin{eqnarray}
&& A(L) \leqslant \int (d^4 p)^L \, \left(\frac{e^{-p^2/\Lambda^2}}{p^2} \right)^I \, 
\left(e^{p^2/\Lambda^2} p^2 \right)^V \nonumber \\
&& \hspace{0.85cm}
= \int (dp)^{4 L} \left(\frac{e^{-p^2/\Lambda^2}}{p^2} \right)^{I - V} \nonumber \\
&& \hspace{0.85cm}
= \int (dp)^{4 L} \left(\frac{e^{-p^2/\Lambda^2}}{p^2} \right)^{L-1} \, .
\label{diver}
\end{eqnarray}
In the last step we used again the topological identity $I = V+L-1$.
The $L$-loops amplitude is UV finite for $L>1$ and it diverges as ``$p^4$" for $L=1$.

Thus only 1-loop divergences exist and the theory is super-renormalizable\footnote{A {\em local} super-renormalizable quantum gravity with a large 
number of metric derivatives was for the first time introduced in \cite{shapiro}.
}.
In these SRQG theories the quantities $\beta$, $\beta_2$, $\beta_0$ and eventually the cosmological constant are renormalized, namely
\be
&& \hspace{-0.7cm}
\mathcal{L}_{\rm Ren} = \mathcal{L}  - 
\sqrt{-g} 
\Big\{  \frac{\beta(Z -1)}{\kappa^2} R  + \lambda (Z_{\lambda} - 1) \nonumber \\
&& \hspace{-0.7cm}
- \beta_2 (Z_2 - 1) (R_{\mu \nu} R^{\mu \nu} - \frac{1}{3} R^2 )
+ \beta_0 ( Z_0 -1 ) R^2 \Big\} \,,
\ee
in which all the coupling must be understood as renormalized at an energy scale $\mu$. On the other hand, the functions $h_i$ are not renormalized because the upper limit $A(L) \leqslant 4$ (\ref{diver}) .

We assume that the theory is renormalized at an energy scale scale $\mu_0$. If we want the bare propagator to possess no other gauge-invariant pole than the transverse physical graviton pole, we have to set  
\be
\hspace{-0.4cm}
\alpha = \beta(\mu_0) \, , \hspace{0.3cm} 
\frac{\alpha_2}{\kappa^2 \Lambda^2} =  \beta_2(\mu_0) \, , \hspace{0.3cm}
\frac{\alpha_0}{6 \kappa^2 \Lambda^2} = \beta_0(\mu_0) .
\label{betaalpha}
\ee
If the energy scale $\mu_0$ is taken as the renormalization point, then $\bar{h}_2 = \bar{h}_0 = \beta(\mu_0) \, \exp(z)$, and only the physical massless spin-2 graviton pole occurs in the bare propagator. In the gauge $\xi = 0$ the propagator in (\ref{fullprop}) reads 
\be 
D_{\mu\nu\rho\sigma}(k) 
= \frac{- i}{(2 \pi)^4} \frac{e^{- k^2/\Lambda^2)}}{\alpha \, ( k^2 + i \epsilon) } 
\Big(  2 P^{(2)}_{\mu \nu \rho \sigma}(k) - 4 P^{(0-s)}_{\mu \nu \rho \sigma}(k) \Big) .\nonumber 
\label{propalpha}
\ee
If we choose another renormalization scale $\mu$, then the bare propagator acquires poles; however, these poles cancel in the dressed physical propagator because 
the renormalization group invariance preserves unitarity in the dressed physical propagator at any energy scale and no other physical pole emerges at any other scale.


\section{Noncommutative spacetime \& Quantum Groups} 

\noindent We unveil in this section the link between one of the SRQG theories analyzed above and the quantum-group structure of spacetime-symmetries proper to noncommuative space-times. The key point is that the two-point function of the super-renormalizable theory can be re-expresed in such a way to exhibit the hidden quantum-group-like structure in the momentum space through the Fourier transform of $\bar{h}_i(- \Box_\Lambda)$ ($i=0,2$). We present in particular two procedures accounting for this result and leading to a particularly simple example of non-commuativity that is well known and has been studied mathematically in depth, namely the $\theta$-Minkowski spacetime with its associated $\theta$-Poincar\'e Hopf algebra of symmetries. We then move to scrutinize possible generalizations within the framework of spacetimes with noncommutativity of the type 
$$[\hat{X}^\mu, \hat{X}^\nu]=i \theta^{\mu \nu}(\hat{X}^\alpha)$$
and conclude with the theorem that, focusing on associative space-time algebras, there is only one possible choice of $\bar{h}_n(\Box_\Lambda)$ compatible with a non-trivial Hopf-algebra structure of space-time symmetries.

\subsection{Emergence of the quantum $\theta$-structure}

\noindent Starting from the expression of the two-point function (\ref{fullprop}), we easily obtain, within an appropriate choice of the gauge, the scalar structure for the graviton propagator to be
\be \label{dampo}
D^{\xi =0}_{\mu\nu\rho\sigma}(k) \sim \frac{- i}{(2 \pi)^4} \frac{e^{-H(k^2/\Lambda^2)}}{k^2 + i \epsilon} \times
{\rm TS} \, , \label{propSR}
\ee
where we start considering $\bar{h}_2(z) = \bar{h}_0(z) := \exp H(z)$ as to be as general as possible and where $H(k^2/\Lambda^2)$ is an entire function of the argument. TS means ``tensorial structure". In order to make explicit the mechanism underlying the result we are going to show, we focus in this first part of the section on a Euclidean \emph{2D} space-time and then consider a phase-space non-commutativity involving $\hat{X}_i$ space-coordinates operators and $\hat{P}_j$ momentum operators characterized by the following Lie-brackets
\be \label{LB}
[\hat{X}^i, \hat{X}^j]= i \theta^{ij} \,,\,\,\,\,\, [\hat{X}^i, \hat{P}_j]= i \delta^i_j \,, \,\,\,\,\, [\hat{P}_i, \hat{P}_j]=0\,,
\ee
namely the Heisemberg non-commutativity between conjugated variables and the Moyal-plane noncommutativity between space-coordinates. It has been shown in Refs.~\cite{Spa, SpaNi} that for a particular choice of $\theta^{ij}$ involving noncommutativity in two of the space-coordinates ({\it e.g.} say $\theta^{3i}=-\theta^{i3}=0$ and $\theta^{ab}=\theta \,\epsilon^{ab}$ with $a,b=1,2$) it is possible to cast the space-noncommutativity on the \emph{2D} plane in terms of Ladder operators and coherent states digonalizing these latter. For instance, assuming $\theta^{ab}=\theta \,\epsilon^{ab}$ one can define $\sqrt{2}\hat{Z}=\hat{X}^1+i \hat{X}^2$ and $\sqrt{2}\hat{Z}^\dagger=\hat{X}^1-i \hat{X}^2$. These new operators fulfill the algebra $[\hat{Z}, \hat{Z}^\dagger]=\theta$, and their eigenstates are labelled as $|z \rangle$ and are such that $\hat{Z}|z \rangle= z |z \rangle$ and $\langle z | \hat{Z}^\dagger = \langle z | \bar{z}$, namely
\be \label{staco}
|z \rangle = \exp(-\frac{z \bar{z}}{\theta})\, \exp({-\frac{z}{\theta} \hat{Z}^\dagger} )|0 \rangle\,.
\ee
These coherent states of the non-commutative plane satisfy the completeness relation $\int dz d\bar{z} | z \rangle \langle z | = \pi \theta$. In quantum field theory the basic non-commutative variables are fields and their conjugated momenta. Coordinates are represented as labels and are commutative. Differently, for a quantum filed theory grounded on (\ref{LB}) we must consider the expectation  value of fields over coherent states (\ref{staco}) encoding space non-commutativity, in order to relate quantization results to standard commutative quantum field theory. This leads to the expansion of quantum fields on a Fourier basis $\langle z |  \exp (i p_j \hat{X}^j) |  z\rangle$ in which such a expectation value is considered, yielding the crucial result
\be \label{primo}
&&\hspace{-0.3cm} \langle z |  \exp (i p_1 \hat{X}^1 + i p_2 \hat{X}^2) |  z\rangle = \\
 &&\hspace{-0.3cm} =\langle z |  \exp (i p_+ \hat{Z}^\dagger)\, \exp (i p_- \hat{Z})\, \exp\left( \frac{p_- p_+}{2}\, [\hat{Z}^\dagger, \hat{Z}] \right)  |  z\rangle\,, \nonumber
\ee
in which the Baker-Campbell-Hausdorff formula has been used (see Appendix \ref{appa}) and the quantities $\sqrt{2}p_\pm=(p_1 \pm i p_2)$ have been defined. Notice also that shrinking to zero the deformation parameter $\theta$ accounts for considering the ``classical limit'' toward standard-commutative quantum filed theory.

We can now generalize this procedure to a non-commutative \emph{4D} space-time and find an energy-momentum exponential-dumping behavior as in (\ref{dampo}), but only if $H(k^2/\Lambda^2)\sim k^2$. We start considering a phase-space involving spacetime coordinates and conjugated momenta of the type
\be \label{LBST}
\hspace{-0.2cm} 
[\hat{X}^\mu, \hat{X}^\nu]= i \theta^{\mu\nu} \,,\,\,\,\,\, [\hat{X}^\mu, \hat{P}_\nu]= i \delta^\mu_\nu \,, \,\,\,\,\, [\hat{P}_\mu, \hat{P}_\nu]=0\,.
\ee
We recall that for $\theta_{\mu 0}\neq 0$, any Lorentzian theory constructed on (\ref{LBST}) is non-unitary \cite{ChaDePreTu}. For the moment we disregard this problem, perform a Wick rotation to the Euclidean spacetime, and show that assuming the only non-zero components $\theta^{03}=-\theta^{30}\equiv \xi\neq 0$ and $\theta^{12}=-\theta^{21} \equiv \theta \neq 0$ it is possible to give sense to a graviton propagator whose scalar structure is expressed by (\ref{dampo}). Let us see here below how it is possible to achieve this result. Together with the Ladder operators $\hat{Z}$ and $\hat{Z}^\dagger$, we consider the choice of $\theta^{\mu \nu}$ specified above and of another class of Ladder operators involving $\hat{X}_1$ and $\hat{X}_3$ coordinates, namely $\sqrt{2}\hat{T}=\hat{X}^0+i X^3$ and $\sqrt{2}\hat{T}^\dagger=\hat{X}^0-i X^3$. It follows that $[\hat{T},\hat{T}^\dagger]=\xi$, and from the type of space-time non-commutativity we assumed above that $[\hat{T},\hat{Z}]=[\hat{T},\hat{Z}^\dagger]=0$, {\it i.e.} the two sectors of Ladder operators can be simultaneously diagonalized. The coherent states for the $\hat{T}$-sector can be constructed in the same way as for the $\hat{Z}$-sector, yielding eigenstates $|t \rangle$ such that $\hat{T}|t \rangle= t |t \rangle$ and $\langle t | \hat{T}^\dagger = \langle t | \bar{t}$, namely
\be \label{staco2}
|t \rangle = \exp(-\frac{t \bar{t}}{\theta})\, \exp({-\frac{t}{\theta} \hat{T}^\dagger} )|0 \rangle\,,
\ee
which is provided with the completeness relation $\int dt d\bar{t} | t \rangle \langle t | = \pi \theta$. We can therefore consider the coherent states $|z,t\rangle=|z\rangle\,|t\rangle$. The relevant formula for expanding quantum fields on a Fourier basis is given by the manipulation of $\langle z,t |  \exp (i p_\mu \hat{X}^\mu) |  z,t\rangle$. This is the expectation value over the coherent state $|z,t\rangle$ of wave-exponentials entering the Fourier-modes expansion of quantum fields on non-commutative space-time, and yields the crucial result
\be \label{cro}
&&\hspace{-0.3cm} 
\langle z,t |  \exp (i p_\mu \hat{X}^\mu) |  z,t\rangle  = \\
 &&\hspace{-0.3cm}  =\langle z |  \exp (i p_+ \hat{Z}^\dagger)\, \exp (i p_- \hat{Z})\, \exp\left( \frac{p_- p_+}{2}\, [\hat{Z}^\dagger, \hat{Z}] \right)  |  z\rangle \times\nonumber \\
 &&\hspace{0.0cm}  \times \langle t |  \exp (i \tilde{p}_- \hat{T}^\dagger)\, \exp (i \tilde{p}_+ \hat{T})\, \exp\left( \frac{\tilde{p}_- \tilde{p}_+}{2}\, [\hat{T}^\dagger, \hat{T}] \right)  |  t\rangle \,. \nonumber
\ee
In (\ref{cro}) we have introduced the quantities $\sqrt{2}\tilde{p}_\pm= (p_0 \pm i p_3)$ and used the Baker-Campbell-Hausdorff formula. Once again we emphasize that, having performed a Wick rotation, our analysis focuses on the Euclidean signature ${\rm sign}(\eta_{\mu\nu})=(+,+,+,+)$, for which problems of unitarity do not appear. Among the super-renormalizable theories previously considered there exist one that can be recast when $\xi=\theta$ as a theory over a non-commutative space-time, with non-commutativity of the type
\be \label{tetap}
[\hat{X}^\mu,\hat{X}^\nu]=i\theta^{\mu \nu}\,.
\ee
Indeed, when  $\xi\!=\!\theta\!=\!1/\Lambda$, the damping factor in (\ref{dampo}), specialized to the case $H(z)=z$, is automatically recreated from the kinematical manipulations reviewed in (\ref{cro}). Moreover, the choice of $\xi=\theta$ preserves Lorentz covariance when we Wick rotate back to the non-commutative space-time with Lorentzian signature. It is well known in literature that the algebra of symmetries for the non-commutative space-time in (\ref{tetap}) is a twisted Hopf algebra $\mathcal{P}_{\theta}$ called $\theta$-Poincar\'e Hopf algebra, and that a theory with a covariant action gives rise in this framework to conserved Noether charges (see {\it e.g.} \cite{Mar_theta}, which extends the work done in \cite{Mar_kappa} for the case of the $\kappa$-Poincar\'e Hopf algebra). Therefore $\theta$-Poincar\'e truly represents an external symmetry (Hopf) algebra, and the same holds for its Lorentz sub-algebra. Moreover, we can construct a covariant theory under the action of the generators of $\mathcal{P}_{\theta}$ that has a new scale invariant, {\it i.e.} $\theta$. 
We will discuss later these implications. We complete the discussion on the emergence of the $\theta$-Poincar\'e symmetry noticing that the tensorial structure in (\ref{fullprop}) does not affect the result of our analysis, as indeed this can be made fully consistent with the $\theta$-Poincar\'e symmetry of this theory \cite{Wess}. Indeed, a remarkable feature of the $\theta$-Poincar\'e quantum groups is that the Lorentz subalgebra is unmodified with respect to the Poincar\'e algebra. This allows to define linear Lorentz transformation and conservation laws following the standard recipe. From now on, we will mention the Lorentz sector, without specifying that it belongs to the $\theta$-Poincar\'e or Poincar\'e algebra Hopf algebra\footnote{Formally, both the $\theta$-Poincar\'e quantum group and the Poincar\'e algebra are Hopf algebra. But the latter one is usually referred to as ``trivial'' Hopf algebra, because of the standard co-commutativity in the co-algebra sector \cite{libriQG}.}.

\subsection{A different philosophy to unveil $\mathcal{P}_\theta$}

\noindent The emergence of the $\theta$-Poincar\'e symmetry-structure does not rely on the particular procedure we adopted in the preceding subsection. For instance, we could have chosen to adopt the ``Weyl system'' procedure \cite{WeBa}, as it has been done in \cite{Mar_theta} at the purpose of analyzing the symmetry-structure of a scalar field theory on $\theta$-Minkowski spacetime. The ``Weyl map'' $\Omega$ associates to any function $f(\hat{x})$ of $\theta$-Minkowski an auxiliary commutative function $f^{(c)}(x)$. The easiest way of implementing this map is to consider the Fourier transform $\tilde{f}(p)$ of $f(\hat{x})$ and then apply on the Fourier modes the Weyl map, namely
\be
f(\hat{x})&=& \Omega\left(f^{(c)}(x)\right)\equiv \Omega \left( \int d^4p \tilde{f}(p) e^{ipx} \right)  \nonumber\\
&=&\int d^4p \tilde{f}(p) \,:e^{i p \hat{x}}:\,,
\ee
in which ``$:\,\cdot\,:$'' denotes an ordering of the non-commutative coordinates associated to $\Omega$. The inverse of the Weyl map, namely $\Omega^{-1}$, is also well defined and is called the Wigner map. Our Wigner map is expressed by the semi-classical limit of the measurement-procedure involving coherent states $|z,t\rangle$, namely
\be \label{equivalenza}
\Omega^{-1}(\cdots)= \langle z,t| \cdots |z,t\rangle\,.
\ee
This procedure also provides a physical picture in our context of the Wigner and Weyl maps. As reminded above, because of the peculiar features of spacetime noncommutativity, a Weyl map selects a particular normal ordering for the spacetime coordinates. Suppose to choose the ``semiclassical-state'' ordering, which is defined in the $x_1-x_2$ plane by the following action of the Weyl map on the Fourier-modes basis elements
\be
&&\Omega|_{x-y}(e^{i p_1x_1 +i p_2 x_2})=  \Omega|_{x-y}(e^{i p_+ \bar{z} + i p_- z })\nonumber\\
&&= e^{i p_+ \hat{Z}^\dagger + i p_- \hat{Z} }= e^{ i p_+ \hat{Z}^\dagger}\, e^{i p_- \hat{Z}} \, e^{ \frac{p_-\! p_+\!}{2}\, [\hat{Z}^\dagger, \hat{Z}]} \,.
\ee
On the whole $\theta$-Minkowski spacetime (\ref{tetap}), in which $\theta_{03}=\theta_{12}=\theta= -\theta_{30}= - \theta_{21}$ and $\theta_{13}=\theta_{23}=\theta_{01}=\theta_{02}=0$, we might consider the definition of ``semiclassical-state'' ordering and the related Weyl map by adopting the coordinates $\hat{T}$ and $\hat{T}^\dagger$,
\be \label{ordo}
&&\hspace{-1cm} 
\Omega\left( e^{i p_\mu x^\mu} \right)= e^{i p_+ \hat{Z}^\dagger}\, e^{i p_- \hat{Z}}\, e^{ \frac{p_- p_+}{2}\, [\hat{Z}^\dagger, \hat{Z}] } \times \nonumber\\
&&\hspace{1cm} \times \,  e^{i \tilde{p}_- \hat{T}^\dagger}\, e^{ i  \tilde{p}_+ \hat{T}}\, e^{\frac{\tilde{p}_- \tilde{p}_+}{2}\, [\hat{T}^\dagger, \hat{T}] } \,.
\ee 
Now define the integration map $\int$ on the non-commutative $\theta$-Minkowski spacetime as the map such that 
\be
\int \Omega(f(x)) \Omega(g(x))=\int d^4 p \tilde{f}(p) \tilde{g}(-p)\, e^{-\theta (p_\mu p^\mu)^2}\,.
\ee
A quantum theory of non commutative fields can now be constructed following the same steps as in \cite{Bala}, namely considering an expansion of the quantum field, fulfilling a Lorentz covariant equation of motion, on a non-commutative Fourier basis 
\be \label{grounded}
&& \!\Psi_r(\hat{X}) = \!\!\int \! d \mu_p \!\left[ a_p \, \Omega \left(  e^{-i p_\mu x^\mu}  \right) \!+\! a^\dagger_p \, \Omega \left(  e^{i p_\mu x^\mu}  \right) \right] 
   \nonumber \\
&& \hspace{0.0cm} 
= \!\!\int \!\frac{d^3\vec{p}}{2 p_0} \!\left[ a_{\vec{p}} \, \Omega \left(  e^{-i p_\mu x^\mu}  \right) \!+\! a^\dagger_{\vec{p}} \, \Omega \left(  e^{i p_\mu x^\mu}  \right) \right] \,  ,
\ee
and then imposing braiding relations on the ladder operators by means of the bi-algebra twisting element 
$$\mathcal{F}_\theta=\exp(\frac{1}{2} \theta^{\mu \nu} P_\mu \otimes P_\nu).$$ 
Notice that in (\ref{grounded}) we have introduced the notation for the Lorentz invariant measure $d\mu_p= d^4 p \,\delta(p^2)=d^3 \vec{p}/2 p_0$. The requirement of compatibility of the covariant action of symmetries on tensor product of states with the tensor product of state on which symmetries have already acted, yields indeed the braiding in the multi-particle states. This peculiar feature of non-commutative quantum field theory enjoying $\theta$-Poincar\'e symmetries are originated by the action of the twisting element, which we define here by means of $\mathcal{F}_\theta\,\triangleright (|p\rangle \otimes | q \rangle) = \mathcal{F}_\theta(p,q)\,|p\rangle \otimes | q \rangle $, {\it i.e.} through
\be \label{com}
a_p a_q\!\!&=&\!\!\mathcal{F}_{\theta}^{-2}(q,p)a_q a_p\,, \nonumber\\
a_p a^\dagger_q \!\!&=&\!\! \mathcal{F}_\theta^{-2}(-q,p) a^\dagger_q a_p + 2 p_0 \delta^4(p-q) \,, \nonumber\\
a^\dagger_p a^\dagger_q\!\!&=&\!\! \mathcal{F}^{-2}_\theta\, a^\dagger_q a^\dagger_p\,.
\ee
The vacuum state of the Fock space is defined by $a_p|0\rangle=0$, and states of the Hilbert by $|p\rangle=a^\dagger_p|0\rangle$. The second quantization procedure hence defined can be applied to the geometric two-tensor field, as defined in \cite{Wess} and perturbative-expanded as in \cite{Modesto}. The second relation in (\ref{com}) is what we need in order to compute the graviton propagator in the non-commutative theory, namely 
\be \label{ulti}
&& D_{\mu\nu\rho\sigma} (\hat{X}^\alpha\!-\!\hat{Y}^\alpha) = \int d\mu_p d\mu_k \nonumber \\
&& \Big \{ \Big[ \langle 0|  \Big( a_p a^\dagger_k \, \Omega(e^{-i p_\mu x^\mu}) \Omega(e^{i k_\mu y^\mu})  \nonumber \\
&& + a^\dagger_p a_k \Omega(e^{i p_\mu x^\mu}) \Omega(e^{-i k_\mu y^\mu})  \Big)  |0\rangle \,\theta(\hat{X}_0-\hat{Y}_0)\Big] + \nonumber \\
&& \Big[ \, \hat{X}_0 \leftrightarrow \hat{Y}_0 \,\,\,
{\rm and}\,\,\, x \leftrightarrow y \,\,\,  \Big] \Big\}  \times  {\rm TS} \nonumber \\
&& = \int \!\! d \mu_p \times {\rm TS} \times  \Omega \left( e^{ip_{\mu}(x^\mu-y^\mu)}  \right) 
\times {\rm pole\,\, structure} \,. \nonumber 
\ee
We emphasize that $D_{\mu\nu\rho\sigma}(\hat{X}^\alpha\!-\!\hat{Y}^\alpha)$ is differs from the expectation value (on the coherent states $|z,t\rangle$) of the propagator of the quantum theory, namely $D_{\mu\nu\rho\sigma} (x^\alpha\!-\!y^\alpha)$. We would have obtained $D_{\mu\nu\rho\sigma} (x^\alpha\!-\!y^\alpha)$ if we had followed the same strategy as in Ref.~\cite{Spa, SpaNi}. In our notation, in terms of the Wigner map, this accounts for
$$\Omega^{-1}\left(D_{\mu\nu\rho\sigma} (\hat{X}^\alpha\!-\!\hat{Y}^\alpha)\right) = D_{\mu\nu\rho\sigma} (x^\alpha\!-\!y^\alpha)$$
that leads to the graviton propagator 
\be \label{ma}
&& D_{\mu\nu\rho\sigma} (x^\alpha - y^\alpha) = \int d\mu_p d\mu_k \\
&& \times
 \Big\{\Omega^{-1} \Big[ \langle 0|  \Big( \!a_p a^\dagger_k \Omega(e^{-i p_\mu x^\mu}) \times
\Omega(e^{i k_\mu y^\mu}) \nonumber \\
&& + a^\dagger_p a_k \Omega(e^{i p_\mu x^\mu})
\Omega(e^{-i k_\mu y^\mu})  \Big) \!|0\rangle \,\theta(\hat{X}_0-\hat{Y}_0)\Big]+ \nonumber \\
&& +\Omega^{-1} \Big[ \,\, x \leftrightarrow   y\,\,  \Big] \Big\} \!\times\! {\rm TS} =  \int d^4p \frac{e^{- p^2/\Lambda^2}}{p^{2}} \times {\rm TS}\,. \nonumber 
\ee
The Fourier transform of the graviton propagator sketched in (\ref{ulti}) by using the ordering introduced in (\ref{ordo}) turns out to give the same value determined henceforth at the beginning of this section, in (\ref{propSR}). 

The procedure incorporated in this second section is more general than that one based on the expectation value on coherent states and must be in general considered as distinct. Nevertheless, this reduces to the one exposed in the preceding section whenever we consider (\ref{equivalenza}) as a concrete definition for the Wigner map. 

\subsection{Uniqueness of the link between quantum groups and SRQG and falsifiability of the theory}

\noindent In this section we prove a simple theorem stating the uniqueness of the link between quantum groups and SRQG. Specifically, we prove that the only non trivial Hopf algebra connected to SRQG is the $\theta$-Poincar\'e Hopf algebra, and that this latter selects only one among the many possible theories (namely, the theory defined by the choice $\bar{h}_2(z)=\bar{h}_0(z)=\exp z$, with $z= - \Box/\Lambda^2$) described in \cite{Modesto}. Thus, in what follows we scrutinize the possibility of generalizing results previously exposed to a wider class of non-commutative spacetimes and prove the impossibility of achieving this goal if we decide not to relax the requirement of associativity for the non-commutative space-time algebra. 

The natural place in order to seek for the generalization of previous results is represented by (\ref{primo}) and the implementation within it of the Baker-Campbell-Hausdorff formula and of its inverse formula, the Zassenhaus formula. Suppose indeed to consider in (\ref{dampo}) the integer function to be $H(k^2/\Lambda^2)=c_1k^2/\Lambda^2 + c_2 (k^2/\Lambda^2)^2$. We address the search for a suitable Lie algebra reproducing this structure for $H(- \Box/\Lambda^2)$ in terms of generic functions $\theta_{34}$, depending\footnote{We recall that the BCH formula and its inverse have been developed considering only Lie-algebra cases. Thus $\theta_{34}$ and $\theta_{12}$ could be rigorously expanded only up to linear order in the generators of the algebra.} on $\hat{Z}$ and $\hat{Z}^\dagger$, and $\theta_{12}$ depending on $\hat{T}$ and $\hat{T}^\dagger$. Commutation relations for the Ladder operators now read
\be \label{ZT}
&&[\hat{Z},\hat{Z}^\dagger]\,=\,\theta_{34}( \hat{Z}, \hat{Z}^\dagger)\,, \nonumber\\
&&[\hat{T},\hat{T}^\dagger]\,=\,\theta_{12} (\hat{T}, \hat{T}^\dagger) \,.
\ee
Maintaining unchanged the definition of $\hat{Z}$ and $\hat{T}$, formulas (\ref{ZT}) yields a space-time non-commutativity of the form
\be \label{luza}
&& [\hat{X}^1,\hat{X}^2] \,=\, i\tilde{\theta}_{12} (\hat{X}_1\,,...\hat{X}_4) \,, \nonumber\\
&& [\hat{X}^3,\hat{X}^4] \,=\, i\tilde{\theta}_{34}(\hat{X}_1\,,...\hat{X}_4) \,.
\ee
Notice that in general both the $\theta$-Minkowski type and $\kappa$-Minkowski \cite{kappa_Min} type of non-commutativity are present in the expansion of the functions $\tilde{\theta}_{12}$ and $\tilde{\theta}_{34}$. Such a co-presence of space-time non-commutativities has been considered in literature \cite{Lukier} in light of its relation with string-theory scenarios\footnote{Expansion of formulae (\ref{luza}) make sense up to second order in a scale $\kappa$ having dimension of inverse energy. Following dimensional arguments, for a spacetime non commutativity of the type 
$$[\hat{X}^\mu,\hat{X}^\nu]=i\theta^{\mu\nu}(\hat{X}^\alpha)$$ 
the only class of deformations of spacetime having classical limit for $\kappa\rightarrow 0$ are of the type 
$$[\hat{X}^\mu,\hat{X}^\nu]=i \kappa^2\, \theta^{\mu\nu}_0 +  i \kappa\, \theta^{\mu\nu}_{(1)\,\,\rho} \hat{X}^\rho+\theta^{\mu\nu}_{(1)\,\,\rho \sigma} \hat{X}^\rho \hat{X}^\sigma,$$
 with $\theta^{\mu\nu}_{(0)}$ and $\theta^{\mu\nu}_{(1)\,\,\rho}$ and $\theta^{\mu\nu}_{(1)\,\,\rho \sigma}$ dimensionless quantities. Within this class of deformations of spacetime, previous expansion describes (for open string first-quantized in \emph{D}$=10$) noncommutative coordinates on \emph{D}-branes providing the localizations of the ends of the strings}. But (\ref{luza}) is not sufficient in order to ensure the desired behavior for the Fourier transform of the integer function $H(k^2/\Lambda^2)$ appearing in the graviton propagator calculation. In other words, we can not achieve the Fourer transform 
\be \label{antonardo}
H(k^2/\Lambda^2)=c_1k^2/\Lambda^2 + c_2 (k^2/\Lambda^2)^2
\ee
on a perturbed background of the form (\ref{peba}) if we still require the non-commutative algebra to be associative. We can prove this theorem considering that two requirements should be fulfilled as necessary conditions in order to add a term like $c_2 (k^2/\Lambda^2)^2$ in 
(\ref{dampo}). The first one reads
\be \label{cca}
\frac{\partial\tilde{\theta}_{12}}{\partial \hat{Z}}=\frac{\partial\tilde{\theta}_{12}}{\partial \hat{Z}^\dagger}=\frac{\partial\tilde{\theta}_{34}}{\partial \hat{T}}=\frac{\partial\tilde{\theta}_{34}}{\partial \hat{T}^\dagger}=0
\ee
and ensures that momenta are not redefined at linear order in $\sqrt{\theta}$, {\it i.e.} for linear Planck mass corrections. The second condition is 
\be \label{ccb}
\frac{\partial^2 \tilde{\theta}_{12}}{\partial \hat{Z} \partial \hat{Z}^\dagger}=\frac{\partial\tilde{\theta}_{12}}{\partial \hat{Z}^\dagger \partial \hat{Z} }=\frac{\partial^2 \tilde{\theta}_{34}}{\partial \hat{T}^\dagger \partial \hat{T}}=\frac{\partial\tilde{\theta}_{34}}{  \partial \hat{T} \partial \hat{T}^\dagger}=\theta
\ee
and ensures the existence of two terms summing in $H(p^2/\Lambda^2)$ within (\ref{dampo}) that are $\theta (p_1^2+p_2^2)^2$ from the $\hat{T}-\hat{T}^\dagger$ sector and $\theta (p_3^2+p_4^2)^2$ from the $\hat{Z}-\hat{Z}^\dagger$ sector. But once summed, these contributions are not sufficient in recreating a covariant $(p^2)^2$ term, which in stead would come from the Fourier transform of $\Box^2$ as it appears in the second term of (\ref{antonardo}). Therefore we should consider now interaction between the two sectors, $\hat{T}-\hat{T}^\dagger$ and $\hat{Z}-\hat{Z}^\dagger$, which would now determine the appearance of mixed terms in $p_2^2$ and $p_3^2$, from one side, and $p_4^2$ and $p_1^2$ from the other side. In the euclidean spacetime, we now label operators within the $\hat{Z}-\hat{Z}^\dagger$ sector as 
\be
\hat{Z}=\hat{Z}_{34}\,, \qquad \hat{Z}^\dagger=\hat{Z}_{34}^\dagger\,,
\ee
and momenta as 
\be
p_+=p_{34}\,, \qquad p_-=p^*_{34}\,. 
\ee
In the $\hat{T}-\hat{T}^\dagger$ sector, operators are now labeled as 
\be
\hat{T}=\hat{Z}_{12}\,, \qquad \hat{T}^\dagger=\hat{Z}_{12}^\dagger\,,
\ee
while momenta are labeled as follows 
\be
\tilde{p}_+=p_{12}\,, \qquad \tilde{p}_-=p^*_{12}\,.
\ee
We emphasize that, in order to obtain a dumping exponential phase-term $\exp\,[-2\theta^2 (p_1^2+p_2^2)(p_3^2+p_4^2)]$ multiplying the other dumping phase-term $\exp\,\{-\theta^2\,[ (p_1^2+p_2^2)^2+(p_3^2+p_4^2)^2]\}$, and hence recreating a covariant exponential dumping phase-factor $\exp - \theta^2 (p^2)^2$, new conditions must be fulfilled about the non-commutativity in the $\hat{X}_1-\hat{X}_3$ plane and in the $\hat{X}_1-\hat{X}_4$ plane, as well as in the  $\hat{X}_2-\hat{X}_3$ and $\hat{X}_2-\hat{X}_4$ planes. These can be derived looking at the exponential 
\be
&&\hspace{-0.5cm} 
\exp i(\hat{X}_1p_1+\hat{X}_2p_2+\hat{X}_3p_3+\hat{X}_4p_4)\nonumber\\
&& \hspace{-0.5cm} 
=\exp i(\hat{Z}_{34}^\dagger p^*_{34} +\hat{Z}_{34} p_{34} +\hat{Z}_{12}^\dagger p^*_{12}+\hat{Z}_{12} p_{12})\,,
\ee
and at its decomposition by means of the Zassenhaus formula, and then imposing that
\be
&&e^{i(p_1\hat{X}_{1}\,+\,p_2\hat{X}_{2}\,+p_3\hat{X}_{3}\,+p_4\hat{X}_{4})}  \\
&& = e^{i(p_1\hat{X}_{1}\,+\,p_2\hat{X}_{2})}\,e^{i(p_3\hat{X}_{3}\,+p_4\hat{X}_{4})} \times  
\nonumber \\
&& \times \, e^{\frac{1}{2}\Gamma[p_1,\, p_2,\,p_3,\, p_4]}\,e^{-\frac{1}{6}\{2 \Phi[p_1,\, p_2,\,p_3,\, p_4] + \Sigma[p_1,\, p_2,\,p_3,\, p_4] \}}\times \nonumber\\
&& \times \, e^{-\frac{1}{24}\{ \Xi[p_1,\, p_2,\,p_3,\, p_4] +3 \Psi[p_1,\, p_2,\,p_3,\, p_4] + 3  \Theta[p_1,\, p_2,\,p_3,\, p_4] \}} \,  ,
 \nonumber 
\ee
with
{\small
\be
&& \hspace{-0.2cm} 
\Gamma[p_1,\, p_2,\,p_3,\, p_4]=[(p_1\hat{X}_{1}\,+\,p_2\hat{X}_{2})\,,\, (p_3\hat{X}_{3}\,+p_4\hat{X}_{4})] \,,   \nonumber\\
&& \hspace{-0.2cm}
\Phi[p_1,\, p_2,\,p_3,\, p_4]=  [(p_3\hat{X}_{3}\,+p_4\hat{X}_{4})\,, \, [(p_1\hat{X}_{1}\,+\,p_2\hat{X}_{2}) \, ,\,
\nonumber \\
&& \hspace{5.35cm}(p_3\hat{X}_{3}\,+p_4\hat{X}_{4})]]       \,    ,   \nonumber\\
&& \hspace{-0.2cm}
\Sigma[p_1,\, p_2,\,p_3,\, p_4] 
=  [(p_1\hat{X}_{1}\,+\,p_2\hat{X}_{2})\,,\,[(p_1\hat{X}_{1}\,+\,p_2\hat{X}_{2}) , \nonumber \\
&& \hspace{5.35cm} (p_3\hat{X}_{3}\,+p_4\hat{X}_{4})]]\,,\nonumber\\
&& \hspace{-0.2cm} 
\Xi[p_1,\, p_2,\,p_3,\, p_4]= [ [ [ (p_1\hat{X}_{1}\,+\,p_2\hat{X}_{2})\,,(p_3\hat{X}_{3}\,+p_4\hat{X}_{4})]\,, 
\nonumber \\
&& \hspace{2.85cm} 
(p_1\hat{X}_{1}\,+\,p_2\hat{X}_{2})]\,,(p_1\hat{X}_{1}\,+\,p_2\hat{X}_{2})]\,,   \nonumber\\
&& \hspace{-0.2cm}
\Psi[p_1,\, p_2,\,p_3,\, p_4]=  [ [ [ (p_1\hat{X}_{1}\,+\,p_2\hat{X}_{2})\,, (p_3\hat{X}_{3}\,+p_4\hat{X}_{4})] \,,  
\nonumber \\
&& \hspace{2.9cm} 
(p_1\hat{X}_{1}\,+\,p_2\hat{X}_{2})]\, , (p_3\hat{X}_{3}\,+p_4\hat{X}_{4})] \,,    \nonumber\\ 
&& \hspace{-0.2cm}
\Theta[p_1,\, p_2,\,p_3,\, p_4]=  [ [ [  (p_1\hat{X}_{1}\,+\,p_2\hat{X}_{2})\,, (p_3\hat{X}_{3}\,+p_4\hat{X}_{4})]\,, 
\nonumber \\
&& \hspace{3cm} 
(p_3\hat{X}_{3}\,+p_4\hat{X}_{4})]\,, (p_3\hat{X}_{3}\,+p_4\hat{X}_{4})]\,. \nonumber 
\ee
}
For arbitrary values of $p_\mu$, the requirement on the Lorentz-invariance of the algebraic sector (and thus on the Lorentz-invariance of the Fourier space) implies from enforcing $\Gamma[p_1,\, p_2,\,p_3,\, p_4]=0$ that 
\be \label{nnc}
[\hat{X}_1,\,\hat{X}_3]=[\hat{X}_1,\,\hat{X}_4]=[\hat{X}_2,\,\hat{X}_3]=[\hat{X}_2,\,\hat{X}_4]=0\,.
\ee
Namely, Lorentz invariance requires that the only type of affordable non-commutativity is the one we considered above on the $\hat{X}_1-\hat{X}_2$ plane and on the $\hat{X}_3-\hat{X}_4$ plane. We would have reached the same conclusion from (\ref{nnc}) by imposing Lorentz-invariance on the Fourier space and hence the simultaneous vanishing of $\Phi[p_1,\, p_2,\,p_3,\, p_4]$ and $\Sigma[p_1,\, p_2,\,p_3,\, p_4]$. Notice that also (\ref{cca}) and (\ref{ccb}), once expressed in terms of $\hat{X}^\mu$ operator, would lead to the same type of inconsistencies. Finally, requirements in (\ref{nnc}) impose the vanishing of both $ \Xi[p_1,\, p_2,\,p_3,\, p_4]$, $\Psi[p_1,\, p_2,\,p_3,\, p_4]$ and $ \Theta[p_1,\, p_2,\,p_3,\, p_4]$. As a consequence, it is manifest the impossibility of recovering a term which goes like $(k^2/\Lambda)^2$  for $H(k^2/\Lambda^2)$ in (\ref{dampo}), if we start from a Lie algebra type of non-commutativity. This result could be in part anticipated. We know indeed that only the twisted $\theta$-Poincar\'e Hopf algebra at the same time preserves Lorentz symmetry, at least in the algebraic sector and in the Fourier space, and consistently realizes the associativity in the module algebra (of space-time coordinates functions).

Finally, the argument developed here above and based on the choice of the particular SRQG theory defined by (\ref{antonardo}), can be repeated for any entire function $H(z)$, thus for any generic SRQG theory. This ends our proof about the uniqueness of the link between SRQG and non-trivial Hopf algebra, specifically the $\theta$-Poincar\'e quantum group. It is not overwhelming to emphasize that the uniqueness, {\it i.e.} having fixed $H(z)$ to a unique function, traces back to the definition of a unique theory of SRQG among the many allowed in the framework of \cite{Modesto}. We also emphasize that such a result relies on the construction of the phase-space with Heisenberg-type of non-commutativity between spacetime and momenta, thus it is consistent with the associativity of the space-time coordinates considered in previous sections\footnote{A notable example of nonassociative spacetime is the Snyder noncommutative space-time (see {\it e.g.} Ref.~\cite{BatMel} and references therein).}. Extending the analysis to deformed phase-space, and hence to a non-associative spacetimes, would not have allowed us to conclude with the same statement.

\subsection{SRQG and Non-commutative Gravity}

\noindent In preparation for the conclusions, we want to address in this section a brief comparison of the model above with the theory of non-commutative geometry and gravity developed in \cite{Wess}, and with seminal works on the relation between differential calculi over a given noncommutative associative algebra and space-time metrics addressed in \cite{Madore1996bb, Cerchiai2000qu}. We first emphasize the differences between the model presented in this letter and the works in \cite{Madore1996bb, Cerchiai2000qu} and \cite{Wess}, and then conclude with a list of points to be investigated in forthcoming works in order to gain a clearer physical picture.

The most striking point we are confronted with is the in-equivalence of our model with the ones addressed in \cite{Madore1996bb, Cerchiai2000qu} and \cite{Wess}. This feature indeed is already evident at the level of the linearized equation of motion for the SRQG theories described in \cite{BM, Modesto}. A first heuristic analysis based on the work reported in references \cite{Barvi} reveals indeed that linearized equation for the model here treated would involve a non-local operator $H_{\mu \nu}^{\,\,\,\,\rho \sigma}(\nabla_\alpha)$ acting on the Ricci scalar $R$ and the Ricci tensor $R_{\mu\nu}$ in the form 
\be \label{GH}
G_{\mu\nu}+ \kappa^2 H_{\mu \nu}^{\,\,\,\,\rho \sigma}(\nabla_\alpha) R_{\rho \sigma}=0\,,
\ee
in which $G_{\mu \nu}$ denotes the Einstein tensor and $\kappa^2\!=\!8 \pi G$ in natural units. In (\ref{GH}) $H_{\mu \nu}^{\,\,\,\,\rho \sigma}(\nabla_\alpha)$ acts as a total derivative only on the Ricci tensor and therefore the associativity condition of an eventual star-produc would not be satisfied. Thus it would be completely meaningless even try to make sense of $H_{\mu \nu}^{\,\,\,\,\rho \sigma}(\nabla_\alpha)$ in terms of a star-product and of an underlying non-commutative of space-time, even in situation in which the background has been fixed and gravity has been linearized at the first order, as for instance when considering $g_{\mu\nu}= \eta_{\mu\nu}+ \kappa h_{\mu \nu}$. Moreover is matter of fact that equation (\ref{GH}) differs from the equation of motion derived in the model studied in \cite{Wess}, where the non-commutative Einstein equations read
\be \label{NCE}
{\rm \bf Ric} -\frac{1}{2} {\bf g} \star \mathfrak{R}=0\,.
\ee
In (\ref{NCE}) the $\star$ product is defined in order to be consistent with the twist element that modifies the diffeomorphism algebra. The non-commutative metric tensor reads locally as ${\bf g}=\theta^j \otimes_\star \theta^i \star g_{ij}$, with $\theta^i$ basis one-forms and $\otimes_\star$ the associative $\star$-tensor product associated to the deformed algebra of non-commutatve tensor fields. The Ricci tensor ${\rm \bf  Ric}$ and the Ricci scalar $\mathfrak{R}$ are derived as the contraction of the Curvature tensor, that is defined in terms of the $\star$-covariant derivative $\nabla^\star_u$ (along any vector field $u$ of the module algebra). In brief, the equation (\ref{NCE}) would read as a modification of the only Einstein tensor, whilst the equation of motion of the SRQG model we have studied involve higher derivatives applied to the square of the curvature tensor and their contractions. However, as this heuristic argument does not provide a solid proof, in order to be confident about the in-equivalence of the two models it would be appropriate to analyze some particular symmetry-reduced solutions, which we will do in the future providing accurate equation of motions in the general curved case.

Another point of difference we should single out is the absence in our framework of a consistent interpretation of the non-locality within the action (\ref{theory}) in terms of a twisted star-product. Following for instance a common procedure (see {\it e.g.} Ref.~\cite{Freidel:2006gc} and references therein), we can express any field theory on non-commutative space-time as a non-local field theory on a commutative space-time, provided that non-locality is described in terms of a star-product. Thus in principle we can ask whether it is possible to do the converse in our framework, recovering a star-product. But we should also consider a twist-element which leave undeformed the Lorentz sector of the Poincar\'e algebra, because of the particular dependence on the D'Alambertian covariant operator $\Box$ in the non-local function $F$. This feature represents a strong constraint for the theories studied in \cite{Modesto}. A twist element would naturally achieve this goal, but it is quite easy to see that from the particular form of $F$ we would not be able to derive the associativity of the star-product, neither the normalization condition for it (see {\it e.g.} section II of Ref.~\cite{Wess}), both of them necessary requirements to recover a twist-element. Therefore we would be naturally lead to search for a generalization of our framework, and more in general of the theories presented in \cite{Modesto}, in order to account for a consistent twist-element. We emphasize that in this latter theoretical framework we would be able to address interesting conceptual questions. Indeed, although in the seminal works in \cite{Madore1996bb, Cerchiai2000qu} cases in which non-commutativity singled out a preferred metric were considered, in \cite{Wess} any moving frame has been treated on equal footing and it has been shown that there are infinitely many metrics compatible with a given non-commutative differential geometry. Moreover, as a consequence of the bicovariant differential calculus and of the framework single out in \cite{Wess}, torsion appears also in the vacuum. This scheme hence implies a deformation of the geodesic motion, and consequences for the equivalence principle should be also investigated in detail. Conversely, the compatibility of the metric and the validity of the equivalence principle are imposed from the beginning in \cite{BM, Modesto} and not quested.

\section{Conclusions}

\noindent Moving from the work in \cite{Modesto} defining a class of SRQG theories, we have shown in this paper that is possible to define a unique SRQG theory provided with a nontrivial Hopf algebra of space-time symmetries. The associated phase space is Heisenberg type, and associativity must be preserved in the non-commutative theory. Specifically, the non-trivial Hopf algebra connected to SRQG is the twisted algebra of $\theta$-Poincar\'e, which shows as a remarkable feature that one of having a Poincar\'e algebra, and thus a Lorentz subalgebra, that are unmodified in the dimension-full parameter $\theta$. For $\theta$-Poincar\'e, deformation emerges in the co-algebra structure and in the other mathematical structures defining the concept of Hopf-algebra, which is a bi-algebra fulfilling certain consistency relations \cite{libriQG}. Therefore, Lorentz transformation and Lorentz covariance is defined in the standard way in this quantum-group symmetric SRQG theory, and locally it makes still sense to say that the Lorentzian theory is invariant under action of the generators of $\mathfrak{so}(3,1)$. Anyway, the presence of $\theta\sim 1/\Lambda^2$ modifications in the co-algebraic sector induces a treatment of many-particle states at the quantum level which opens the path to entanglement effects. Moreover, the fuzziness of spacetime might have now consequences on the principe of equivalence in the quantum theory. These are suggestive questions that can be addressed only at the level of a full quantized theory, we leave therefore them for future developments.

\appendix 

\section{$P^{(2)}, P^{(1)},  P^{(0-s)},  P^{(0-s\omega)}$-Tensors} \label{P}

\noindent We furnish here below the expression for some quantities introduced in Section \ref{due}, namely 
\be
 && P^{(2)}_{\mu \nu \rho \sigma}(k) = \frac{1}{2} ( \theta_{\mu \rho} \theta_{\nu \sigma} +
 \theta_{\mu \sigma} \theta_{\nu \rho} ) - \frac{1}{3} \theta_{\mu \nu} \theta_{\rho \sigma}  ,
 \nonumber \\
 && P^{(1)}_{\mu \nu \rho \sigma}(k) = \frac{1}{2} \left( \theta_{\mu \rho} \omega_{\nu \sigma} +
 \theta_{\mu \sigma} \omega_{\nu \rho}  + 
 \theta_{\nu \rho} \omega_{\mu \sigma}  +
  \theta_{\nu \sigma} \omega_{\mu \rho}  \right) ,
 \nonumber \\
 && P^{(0 - s)} _{\mu\nu\rho\sigma} (k) = \frac{1}{3}  \theta_{\mu \nu} \theta_{\rho \sigma} \,\, , \hspace{0.1cm}
P^{(0 - \omega)} _{\mu\nu\rho\sigma} (k) =  \omega_{\mu \nu} \omega_{\rho \sigma}, \nonumber \\
&& P^{(0 - s \omega)} _{\mu\nu\rho\sigma}  = \frac{1}{\sqrt{3}}  \theta_{\mu \nu} \omega_{\rho \sigma}, 
\,\,
 \hspace{0.1cm} P^{(0 - \omega s )} _{\mu\nu\rho\sigma} =  \frac{1}{\sqrt{3}} \omega_{\mu \nu} \theta_{\rho \sigma}, \nonumber \\ 
 &&\theta_{\mu \nu} = \eta_{\mu \nu} - \frac{k_{\mu} k_{\nu}}{k^2} \,\, , \hspace{0.1cm}
 \omega_{\mu \nu} = \frac{k_{\mu} k_{\nu}}{k^2}. 
\ee

\section{BCH formulae} \label{appa}

\noindent We summarize in this appendix some useful formulae that we have used in the above sections. We first consider a linear operator $A$, which is defined by means of
\vspace{0.8cm}
\begin{equation}
\exp{A} := \sum_{k=0}^{\infty} \frac{1}{k!}A^k.
\end{equation}
As a consequence, 
$\partial_{\tau} e^{\tau A} = A e^{\tau A}= e^{\tau A} A$.
Let us consider another linear operator $B$, and let be $B(\tau)=e^{\tau A} B e^{-\tau A}$. The {\sl Sophus-Lie} formula then provide us with the following series representation for $B(\tau)$:
\begin{equation}
B(\tau)= \sum_{m=0}^{\infty} \frac{{\tau}^m}{m!}B_m,
\end{equation}
in which $B_m =[A,B]_m := [A,[A,B]_{m-1}]$ and $B_0:=B$.
The
{\sl Baker-Campbell-Hausdorff (BCH)} formula is a particular case of the {\sl Sophus-Lie} formula. Setting $\tau =1$, one obtains indeed
\begin{equation}
e^A \; B e^{-A} = \sum_{m=0}^{\infty} \frac{1}{m!} B_m.
\end{equation}
This latter expression can be re-manipulated in the form
\begin{equation}
[B,e^{-A}] = e^{-A} \left( [A,B]+\frac{1}{2} [A,[A,B]] + \cdots
\right),
\end{equation}
or
\begin{equation}
[e^A,B] = \left( [A,B]+\frac{1}{2} [A,[A,B]] + \cdots
\right) e^A.
\end{equation}
Furthermore, in addition to the BCH formula, there is another expression which is also referred to as BCH formula, but which is due to Eugene Dynkin. This latter expression provides us with the multiplication law for two exponentials of linear operators within the assumptions $[A,[A,B]] = [B,[B,A]] = 0$, corresponding to a central algebra in our $\theta$-Minkowski case. It follows that
\begin{equation}
e^A e^B = e^{A+B} e^{\frac{1}{2}[A,B]}\,,
\end{equation}
and reshuffling this latter expression, one obtain the Zassenhaus formula at the second order
\be
e^{A+B} = e^A \, e^B\, e^{- \frac{1}{2}[A,B]}\,.
\ee
As for practical reasons we were mostly interested to the Zassenhaus formula up to the fourth order, here below we furnish it for completeness
\be
&& \hspace{-1cm} e^{A+B} = e^A \, e^B\, e^{-\frac{1}{2}[A,B]} \, e^{\frac{1}{6}\{ 2 [B,[A, B]] + [A, [A, B]]   \}}\times\nonumber\\
&&\hspace{-1cm} \times e^{-\frac{1}{4}\{ [[[A,B],A], A] + 3 [[[A, B], A], Y] + 3 [[[A, B], B], B]  \}}\,.
\ee

\begin{acknowledgments}
\noindent S.A. and A.M. acknowledge support from NSF CAREER grant. L.M.  acknowledges support to his research at Perimeter Institute by the Government of Canada through Industry Canada and by the Province of Ontario through the Ministry of Research \& Innovation.
\end{acknowledgments}


\vspace{-0.22cm}

\begin{thebibliography}{99}

\bibitem{BM} T. Biswas, E. Gerwick, T. Koivisto, A. Mazumdar, 
accepted in Phys. Rev. Letters [arXiv:1110.5249v2].

\bibitem{Modesto}  L.~Modesto,
  arXiv:1107.2403 [hep-th];
L. Modesto [arXiv:1202.0008 [hep-th]].

\bibitem{NL1}
A. S. Koshelev [arXiv:1112.6410 [hep-th]].

\bibitem{NL2}
Alexey S. Koshelev, Sergey Yu. Vernov 
[arXiv:1202.1289 [hep-th]].

\bibitem{NL3}
A. S. Koshelev, AIP Conf. Proc. 1241 (2010) 630-638 
[arXiv:0912.5457 [hep-th]]. 



\bibitem{NS1}
 P.~Nicolini, A.~Smailagic and E.~Spallucci,
 [arXiv:hep-th/0507226].
%
\bibitem{NS2}
 P.~Nicolini, 
 J.\ Phys.\ A  38, L631 (2005)
 [arXiv:hep-th/0507266].
 %
 \bibitem{NS3}
%
 P.~Nicolini, A.~Smailagic and E.~Spallucci,
 Phys.\ Lett.\  B 632, 547 (2006)
 [arXiv:gr-qc/0510112].
%
\bibitem{NS4}
 T.~G.~Rizzo, 
 JHEP 0609, 021 (2006)
 [arXiv:hep-ph/0606051].
%
\bibitem{NS5}
 E.~Spallucci, A.~Smailagic and P.~Nicolini,
 Phys.\ Rev.\  D 73, 084004 (2006)
 [arXiv:hep-th/0604094].
%
\bibitem{NS6}
 S.~Ansoldi, P.~Nicolini, A.~Smailagic and E.~Spallucci, 
 Phys.\ Lett.\  B 645, 261 (2007)
 [arXiv:gr-qc/0612035].
%
\bibitem{NS7}
 E.~Spallucci, A.~Smailagic and P.~Nicolini, 
 Phys.\ Lett.\  B 670, 449 (2009)
 [arXiv:0801.3519 [hep-th]].
%
\bibitem{NS8}
 P.~Nicolini,
 Int.\ J.\ Mod.\ Phys.\  A 24, 1229 (2009)
 [arXiv:0807.1939 [hep-th]].
%
\bibitem{NS9}
 R.~Casadio and P.~Nicolini, 
 JHEP 0811, 072 (2008)
 [arXiv:0809.2471 [hep-th]].
%
\bibitem{NS10}
 I.~Arraut, D.~Batic and M.~Nowakowski, 
 Class.\ Quant.\ Grav.\  26, 245006 (2009)
 [arXiv:0902.3481 [gr-qc]].
%
\bibitem{NS11}
 P.~Nicolini and E.~Spallucci, 
 Class.\ Quant.\ Grav.\  27, 015010 (2010)
 [arXiv:0902.4654 [gr-qc]].
%
\bibitem{NS12}
 A.~Smailagic and E.~Spallucci, 
 Phys.\ Lett.\  B 688, 82 (2010)
 [arXiv:1003.3918 [hep-th]].
%
\bibitem{NS13}
 D.~M.~Gingrich,
 JHEP 1005, 022 (2010)
 [arXiv:1003.1798 [hep-ph]].
%
\bibitem{NS14}
 I.~Arraut, D.~Batic and M.~Nowakowski,
 J.\ Math.\ Phys.\  51, 022503 (2010)
 [arXiv:1001.2226 [gr-qc]].
%
\bibitem{NS15}
 R.~Banerjee, S.~Gangopadhyay and S.~K.~Modak, 
 Phys.\ Lett.\  B 686, 181 (2010)
 [arXiv:0911.2123 [hep-th]].
%
\bibitem{NS16}
 P.~Nicolini, 
 Phys.\ Rev.\  D 82, 044030 (2010)
 [arXiv:1005.2996 [gr-qc]].
 \bibitem{NS17}
L. Modesto, A. Randono,  
[arXiv:1003.1998 [hep-th]].
\bibitem{NS18}
F. Caravelli, L. Modesto, 
Phys. Lett. B 702 (2011) 307-311 
[arXiv:1001.4364 [gr-qc]].
%
\bibitem{NS19}
 L.~Modesto and P.~Nicolini, 
 Phys.\ Rev.\  D 82, 104035 (2010)
 [arXiv:1005.5605 [gr-qc]].
%
\bibitem{NS20}
 R.~B.~Mann and P.~Nicolini, 
 Phys. Rev. D 84 (2011) 064014 
 arXiv:1102.5096 [gr-qc].
 \bibitem{NS21}
%
J. R. Mureika and P. Nicolini,  
Phys. Rev. D 84 (2011) 044020 
[arXiv:1104.4120 [gr-qc]].

 

\bibitem{ModestoMoffatNico}
L. Modesto, J. W. Moffat, P. Nicolini, 
Phys. Lett. B 695, 397-400 (2011)
[arXiv:1010.0680 [gr-qc]].

\bibitem{Unpart}
P. Gaete, J. A. Helayel-Neto, E. Spallucci, 
Phys. Lett. B 693, 155-158 (2010) [arXiv:1005.0234 [hep-ph]].


\bibitem{Stelle} 
K.S. Stelle, 
Phys. Rev. D 16, 953-969 (1977). 

\bibitem{Tombo} 
E. T. Tomboulis, 
[hep-th/9702146v1]. 

\bibitem{Spa} A. Smailagic and E. Spallucci, 
J.Phys. A36 (2003) L517-L521, [hep-th/0308193v2].

\bibitem{Shapirobook} I. L. Buchbinder, Sergei D. Odintsov, I. L. Shapiro, 
  ``Effective action in quantum gravity", IOP Publishing Ltd 1992. 


\bibitem{SpaNi} E. Spallucci, A. Smailagic and P. Nicolini,
Phys. Rev. D 73, 084004 (2006) [hep-th/0604094v1]. 

\bibitem{ChaDePreTu} M. Chaichian, A. Demichev, P. Presnajder and A. Tureanu, 
Eur. Phys. J. C20 (2001) 767.

\bibitem{Mar_theta} 
G. Amelino-Camelia, F. Briscese, G. Gubitosi, A. Marciano, P. Martinetti and F. Mercati, 
Phys.Rev. D78 (2008) 025005 [arXiv:0709.4600[hep-th]]; 
G. Amelino-Camelia, G. Gubitosi, A. Marciano, P. Martinetti, F. Mercati, D. Pranzetti and R. A. Tacchi, 
Prog. Theor. Phys. Suppl. 171 (2007) 65-78 [arXiv:0710.1219[gr-qc]]; 
A. Marciano, 
Proceedings of the XII Marcel Grossmann Meeting (Paris 2009) [arXiv:1003.0395 [hep-th]].

\bibitem{Mar_kappa} A.~Agostini, G.~Amelino-Camelia, M.~Arzano, A.~Marciano and R.~A.~Tacchi,
  Mod.\ Phys.\ Lett.\  A {\bf 22}, 1779 (2007)
  [arXiv:hep-th/0607221];
  M.~Arzano and A.~Marciano,
  Phys.\ Rev.\  D {\bf 75}, 081701 (2007)
  [arXiv:hep-th/0701268]; 
  M.~Arzano and A.~Marciano,
  Phys.\ Rev.\  D {\bf 76}, 125005 (2007)
  [arXiv:0707.1329 [hep-th]].

\bibitem{Wess} 
P. Aschieri, C. Blohmann, M. Dimitrijevic, F. Meyer, P. Schupp and J. Wess, 
Class. Quant. Grav. 22 (2005) 3511-3532 [hep-th/0504183]; 
P. Aschieri, M. Dimitrijevic, F. Meyer and J. Wess, 
Class. Quant. Grav. 23 (2006) 1883-1912 [hep-th/0510059].

\bibitem{WeBa} J. Baez, I.E.Segal and Z. Zhou, 
Introduction to Algebraic and constructive Quantum Field Theory, 
Princeton University Press (1992).

\bibitem{Bala} A. P. Balachandran, A. Pinzul and B. Qureshi, arXiv:hep-th/0508151; A. P. Balachandran, T. R. Govindarajan, G. Mangano, A. Pinzul, B. A. Qureshi and S. Vaidya, Phys. Rev. D75, 045009 (2007) 
[hep-th/0608179].

\bibitem{libriQG} V.~Chari and A. Pressley,
 A guide to quantum groups, 
 (Cambridge University Press, Cambridge UK, 1994); 
 S.~Majid, Foundations of Quantum Group Theory, 
 (Cambridge University Press, Cambridge UK, 2000).

\bibitem{kappa_Min} S. Majid and H. Ruegg, Phys. Lett. B 334 (1994) 348 [hep-th/9405107].

\bibitem{Lukier} N. Seiberg and E. Witten, JHEP 09 (1999) 032, hep-th/9908142; Chong-Sun Chu, 
[hep-th/0502167].

\bibitem{VN} P. Van Nieuwenhuizen, 
Nuclear Physics B 60 478-492 (1973).

\bibitem{shapiro}
M. Asorey, J.L. Lopez, I.L. Shapiro,  
Intern. Journal of Mod. Phys. A12, 5711-5734 (1997)
[hep-th/9610006]. 












\bibitem{Madore1996bb} 
  J.~Madore and J.~Mourad,
  J.\ Math.\ Phys.\  {\bf 39}, 423 (1998)
  [gr-qc/9607060].
  
\bibitem{Cerchiai2000qu} 
  B.~L.~Cerchiai, G.~Fiore and J.~Madore,
  Commun.\ Math.\ Phys.\  217, 521 (2001)
  [math/0002007 [math-qa]].

\bibitem{Freidel:2006gc} 
  L.~Freidel, J.~Kowalski-Glikman and S.~Nowak,
  Phys.\ Lett.\ B 648, 70 (2007)
  [hep-th/0612170].

\bibitem{BatMel} 
  M.~V.~Battisti and S.~Meljanac,
  Phys.\ Rev.\ D 82, 024028 (2010)
  [arXiv:1003.2108 [hep-th]].
\bibitem{Barvi} 
A. O. Barvinsky
[arXiv:1107.1463 [hep-th];
A. O. Barvinsky, Phys. Rev. D 71 (2005) 084007 [hep-th/0501093v2]; 
A. O. Barvinsky, Phys. Lett. B 572 (2003) 109-116 [hep-th/0304229v3];
H. W. Hamber, (UC, Irvine) and R. M. Williams, 
Phys. Rev. D72, 044026 (2005) [hep-th/0507017]; 
N. Arkani-Hamed, S. Dimopoulos, G. Dvali, G. Gabadadze,
[hep-th/0209227v1]; 
G. Calcagni, G. Nardelli, 
JHEP 1002, 093 (2010) 
[arXiv:0910.2160 [hep-th]].


\end{thebibliography}
\end{document}